\documentclass[pra,twocolumn,floatfix,superscriptaddress]{revtex4-1}
\usepackage{dcolumn,amsmath}
\usepackage{graphicx}
\usepackage{bm}
\usepackage{amssymb}
\usepackage{hyperref}
\usepackage{multirow}
\usepackage{footnote}

\hypersetup{
    colorlinks,
    citecolor=blue,
    filecolor=blue,
    linkcolor=blue,
    urlcolor=blue
}

\newcommand{\abinitio}{{\it ab~initio} }

\newcommand{\Eeff}{$E_{\rm eff}$}
\newcommand{\eEDM}{{{\it e}EDM}}

\newcommand{\ecm}{\ensuremath{e {\cdotp} {\rm cm}}} 
\newcommand{\kA}{k_{\cal A}}
\newcommand{\HA}{H_{\cal A}}
\newcommand{\Wa}{W_{\rm a}}

\usepackage{xcolor}
\usepackage{pdfcomment}
\begin{document}

%  \title{Radium monofluoride, RaF: a potential search candidate for P- and T,P-odd effects}

\title{{\it Ab initio} study of radium monofluoride, RaF, as a candidate to search for P$-$ and T,P$-$ violation effects}

\author{A. D.\ Kudashov}
\email{kudashovad.@gmail.com}
\author{A. N.\ Petrov}
\author{L. V.\ Skripnikov}
\author{N.~S.\ Mosyagin}
\affiliation{Dept. of Physics, Saint Petersburg State University, Saint Petersburg, Petrodvoretz 198904, Russia}
\affiliation{Petersburg Nuclear Physics Institute, Gatchina, Leningrad district 188300, Russia}
\author{T. A. Isaev}
\affiliation{Clemens-Sch{\"{o}}pf Institute, TU Darmstadt, Petersenstr. 22, 64287 Darmstadt, Germany}
\author{R. Berger}
\affiliation{Clemens-Sch{\"{o}}pf Institute, TU Darmstadt, Petersenstr. 22, 64287 Darmstadt, Germany}
\affiliation{Fachbereich Chemie, Philipps-Universit{\"a}t Marburg, Hans-Meerwein-Str. 4, 35032 Marburg, Germany}
\author{A. V.\ Titov}
\homepage{http://www.qchem.pnpi.spb.ru}
\affiliation{Dept. of Physics, Saint Petersburg State University, Saint Petersburg, Petrodvoretz 198904, Russia}
\affiliation{Petersburg Nuclear Physics Institute, Gatchina, Leningrad district 188300, Russia}

\date{\today}

\begin{abstract}

Relativistic \abinitio calculations have been performed to assess the suitability of RaF for experimental search of P$-$ and T,P$-$violating interactions. The parameters of P$-$ and T,P$-$odd terms of the spin-rotational Hamiltonian have been calculated for the $^{2}\Sigma$ electronic ground state of $^{223}$RaF molecule. They include the $\Wa$ parameter, which is critical in experimental search for nuclear anapole moment and the parameters $W_d$ and $W_{\rm SP}$ required to obtain restrictions on the electric dipole moment of the electron and T,P$-$odd scalar$-$pseudo\-scalar interactions, respectively. The parameter $X$ corresponding to the ``volume effect'' in the T,P$-$odd interaction of the $^{223}$Ra nuclear Schiff moment with electronic shells of RaF has also been computed. Spectroscopic and hyperfine structure constants for $^{223}$RaF and $^{223}$Ra$^{+}$ have been computed as well, demonstrating the accuracy of the methods employed.

\end{abstract}

\maketitle

\section*{Introduction}

Manifestations of interactions which are not symmetric with respect to time (T) or spatial (P) inversions (T,P$-$ and P$-$ odd interactions) are of great interest for modern physics.  Particularly, the observation of an electron electric dipole moment (\eEDM) at a level significantly larger than $10^{-38} \ecm$ would indicate the presence of New Physics beyond the Standard model.  Popular extensions of the Standard model of particle physics predict \eEDM\ magnitude of  $10^{-26}-10^{-29} \ecm$ \cite{Commins:98}.  It was realized many years ago  \cite{Sandars:65, Sandars:67, Shapiro:1968, Labzowsky:78, Sushkov:78, Gorshkov:79} (see also review \cite{Ginges:04} and book \cite{Khriplovich:11}) that very promising experiments towards the search for violation of fundamental symmetries could be performed on atoms, molecules and solids containing heavy elements. Effects connected with parity-violating interactions can be considerably enhanced in such systems. 

However, the enhancement cannot be measured directly in a single experiment and, thus, should be calculated theoretically. Recently, a very strict upper bound on \eEDM~\mbox{($<8.7\cdotp10^{-29}\ecm$)} was obtained in ThO molecular beam experiments \cite{ACME:14a}, based on the computed effective electric field acting on the electron from Refs.~\cite{Skripnikov:13c,Fleig:14}.

Diatomic molecules containing heavy nuclei look very promising and, in fact, turn into the main probe for P$-$ and T,P$-$violating effects in the low-energy sector.

Recently, considerable effort has been put into \abinitio calculations of the nuclear-spin-dependent (NSD) P$-$odd interaction constant $\Wa$ for RaF \cite{Isaev:10, Isaev:12} ($W_A$ in \cite{Borschevsky:13}). Here we present results of \abinitio coupled-cluster calculations of $\Wa$ for $^{223}$RaF, together with other T,P$-$odd term parameters and spectroscopic constants of this radical. Computed hyperfine structure constants (HFS) for $^{223}$Ra$^+$ allow to check the accuracy of the theoretical methods used.

\section{Theoretical background}

Following the designations in \cite{Isaev:12} for consistency with previous works, the term in the spin-rotational Hamiltonian associated with the NSD parity violating effects is 
\begin{equation}
\hat{h}^I_{\rm PV} = \frac{G_\mathrm{F}}{\sqrt{2}}
\sum_{A,i} k_{{\cal A},A}
\bm{\alpha}\cdot\bm{I}_A
\rho_A(\bm{r}_i),
\label{general hamiltonian}
\end{equation}
where $A$ labels all the system's nuclei, $k_{{\cal A},A}$ are the dimensionless strength constants, \mbox{$G_\mathrm{F}~=~2.22249 \cdot 10^{-14}~E_\mathrm{h}\cdot a_0^3$} is the Fermi coupling constant, $E_\mathrm{h}$ is the Hartree atomic unit of energy, $\bm{\alpha}$ is a one-electron operator made up of Dirac matrices
\begin{eqnarray*}
  \bm{\alpha}=
  \left(  
  \left(\begin{array}{cc}
  0      & \sigma_x \\
  \sigma_x & 0 \\
  \end{array}\right),\left(\begin{array}{cc}
  0      & \sigma_y \\
  \sigma_y & 0 \\
  \end{array}\right),\left(\begin{array}{cc}
  0      & \sigma_z \\
  \sigma_z & 0 \\
  \end{array}\right)
  \right)
\end{eqnarray*}
with $\sigma_x,\sigma_y,\sigma_z$ being the $2 \times 2$ Pauli matrices and $0$ a $2 \times 2$ zero matrix, $\bm{r_i}$ is the displacement of unpaired electron $i$ from nucleus $A$, and, finally, $\bm{I}_A$ and $\rho_A(\bm{r})$ are the dimensionless reduced nuclear spin operator and the nuclear density distribution (normalized to unity), respectively. Taking into account the fact that the matrix element of the NSD P$-$odd interactions scales as $\sim Z^2$ \cite{Moskalev:76}, our main concern in this particular case shifts to the NSD parity violating effects due to the radium nucleus. With that in mind, Eq.~(\ref{general hamiltonian}) can be reduced to \cite{Borschevsky:13}
\begin{equation}
\hat{h}^I_{\rm PV} \approx k_{{\cal A},\rm Ra}\frac{G_\mathrm{F}}{\sqrt{2}}
\bm{\alpha}\cdot\bm{I}_{\rm Ra}
\rho_{\rm Ra}(\bm{r}).
\label{Ra hamiltonian}
\end{equation}

For convenience, index Ra is omitted in discussions below ($k_{{\cal A},\rm Ra} \rightarrow \kA$ etc.). It should be noted, that different definitions and designations of $\kA$ are used by different researchers (see Ref.~\cite{Nahrwold:14} and references within). Aside from the nuclear anapole moment, the electroweak neutral coupling between the electron vector and the nucleon axial-vector currents also give rise to the NSD PNC (parity non-conservation) effects \cite{Novikov:77}. However, the nuclear anapole moment contribution to the NSD interaction is expected to be dominant in RaF since it scales as $\sim {\rm A}^{2/3}$, where ${\rm A}$ is the atomic number of Ra \cite{Flambaum:84a}. Finally we have a small contribution to the NSD PNC effects induced by the perturbation of nuclear-spin-independent weak interaction by the hyperfine interaction which also scales as $\sim {\rm A}^{2/3}$
\cite{Gorshkov:82}. 

Following Eqs.~(\ref{general hamiltonian}) and (\ref{Ra hamiltonian}) in case of the $^{2}\Sigma$ electronic ground state of RaF, the NSD parity violating interaction gives rise to a P$-$odd contribution to the effective spin-rotational Hamiltonian, that can be written as \cite{Flambaum:85b,Kozlov:95,DeMille:08}
\begin{equation}
     \HA^\mathrm{eff}= \kA \Wa
\left( \bm{n} \times \bm{S'} \right)\cdot \bm{I},
 \label{s-r hamiltonian}
\end{equation}
where $\bm{S'}$ is the effective electron spin (as defined in Ref. \cite{Kozlov:95}) and $\bm{n}$ is the unit vector directed along the molecular axis from the heavier (Ra) to the lighter (F) nucleus. The electronic parameter $\Wa$ can be written as
\begin{equation}
     \Wa=\frac{G_\mathrm{F}}{\sqrt{2}}
     \left\langle \Psi_{^{2}\Sigma_{1/2}} \left\vert
     \rho(\bm r) {\alpha_+}
     \right\vert \Psi_{^{2}\Sigma_{-1/2}} \right\rangle,
 \label{W_a}
\end{equation} 
where $\Psi$ is the electonic wave function of the considered RaF state and $\alpha_+$ is defined as
\begin{eqnarray*}
  \alpha_+=\alpha_x+\mathrm{i}\alpha_y=
  \left(\begin{array}{cc}
  0      & \sigma_x \\
  \sigma_x & 0 \\
  \end{array}\right)+
  \mathrm{i}\left(\begin{array}{cc}
  0      & \sigma_y \\
  \sigma_y & 0 \\
  \end{array}\right).
\end{eqnarray*}
When $\Wa$ is accurately known from electronic structure calculations, one can determine the isotope specific constant $\kA$ from a successful molecular experiment.

To interpret results of complementary molecular experiments in terms of the \eEDM\ one should know the effective electric field \Eeff\ acting on the electron. In turn, \Eeff\ can be expressed in terms of the T,P$-$odd interaction parameter $W_d$ (discussed in Refs.\ \cite{Kozlov:87, Kozlov:95, Titov:06amin, Baklanov:10}): $E_{\rm eff}=W_d|\Omega|$, where \mbox{$\Omega=\langle\Psi|\bm{J}\cdot\bm{n}|\Psi\rangle$} and $\bm{J}$ is the total electronic angular momentum ($\Omega= \pm 1/2$ for the considered $^{2}\Sigma$ electronic state of RaF). The parameter $W_d$ itself is written as
\begin{equation}
\label{matrelem}
W_d = \frac{1}{\Omega}\langle\Psi_{^{2}\Sigma_{1/2}}|\sum_i\frac{H_d(i)}{d_e}|\Psi_{^{2}\Sigma_{1/2}}\rangle,
\end{equation}
where $d_e$ is the value of \eEDM ,
\begin{eqnarray}
  H_d=2d_e
  \left(\begin{array}{cc}
  0 & 0 \\
  0 & \bm{\sigma} \cdot \bm{E} \\
  \end{array}\right),
 \label{Wd}
\end{eqnarray}
$\bm{E}$ is the inner molecular electric field and $\bm{\sigma}$ is the vector of Pauli spin matrices.

Another T,P$-$odd interaction is the scalar$-$pseudo\-scalar nucleus$-$electron neutral current interaction, which is given by the following operator (see \cite{Ginges:04}):
\begin{eqnarray}
  H_\mathrm{SP}=\mathrm{i}\frac{G_\mathrm{F}}{\sqrt{2}}Zk_\mathrm{SP}\gamma_0\gamma_5\rho(\textbf{r}),
 \label{Hsp}
\end{eqnarray}
where $\gamma_0$ and $\gamma_5$ are the Dirac matrices and $k_\mathrm{SP}$ is a dimensionless coupling constant.

To extract the fundamental $k_\mathrm{SP}$ constant from an experiment one needs to know the factor $W_{\mathrm{SP}}$ (designated $W_{\mathrm{T,P}}$ in Ref.~\cite{Dmitriev:92, Skripnikov:13c} or $W_{\mathrm{s}}$ in Ref.~\cite{Isaev:12}), which is determined by the electronic structure of a studied molecule on a given nucleus:
\begin{equation}
\label{WTP}
W_{\mathrm{SP}} = \frac{1}{\Omega}
\langle \Psi_{^{2}\Sigma_{1/2}}|\sum_i\frac{H_\mathrm{SP}(i)}{k_\mathrm{SP}}|\Psi_{^{2}\Sigma_{1/2}}
\rangle.
\end{equation}

Another experimentally detectable source of T,P-invariance in RaF might be the EDM induced by the Schiff moment of Ra nucleus, $S$ \cite{Sushkov:84a, Flambaum:86}. Due to nuclear octupole deformations \cite{Auerbach:96, Auerbach:97} $S(\rm{Ra})$ should be exceptionally large, e.g., exceeding $S(Tl)$ by $\sim$ 200 times. The most stringent constraint on the atomic EDM to date comes from experiment on $^{199}$Hg \cite{Griffith:09}. According to Ref.~\cite{Dobaczewski:05}, the Schiff moment of $^{225}$Ra may surpass that of $^{199}$Hg by two orders of magnitude, making Ra a very promising candidate for further EDM experiments. The atomic EDM is further enhanced in Ra since the Schiff moment contribution increases faster than $Z^2$. In RaF the observable T,P$-$odd effect associated with the Schiff moment can be expressed in terms of the following effective Hamiltonian \cite{Hinds:80a,Sushkov:84}:
\begin{equation}
     H_{X}= 6 S(\mathrm{Ra}) X \bm{\sigma}_{\mathrm{Ra}} \cdot \bm{n}\ ,
 \label{HX}
\end{equation}
where $\bm{\sigma}_{Ra}$ is the Ra nuclear spin operator and $X$ is determined by the electronic structure of the radical:
\begin{equation}
    X=\frac{2\pi}{3}\left[\frac{\partial}{\partial z} \rho_{e}\left(\bm{r}\right)\right]_{x,y,z=0} 
  \label{X},
\end{equation}
where  $\rho_{e}\left(\bm{r}\right)$ is the electronic density calculated from the four-component wave function $\Psi_{^{2}\Sigma}$.

The constants defined in Eqs. (\ref{W_a}, \ref{matrelem}, \ref{WTP}, \ref{X}) cannot be directly measured experimentally. Their corresponding operators are most sensitive to the wave function of the valence electrons (electron spin density) in the region near the heavy nucleus.  Thus, the standard way to verify the accuracy of the computed electron spin density in the core region (the region near the heavy nucleus) is to calculate the hyperfine structure tensor, which can be measured experimentally (see e.g. \cite{Titov:06amin}). In case of RaF, the tensor has two independent components, which can be written as $A_{\parallel}$ and $A_{\perp}$:
\begin{equation}
 \label{Apar}
A_{\parallel}=\frac{\mu(\rm Ra)}{I\Omega}
   \langle
   \Psi_{^{2}\Sigma_{1/2}}\vert\sum_i\left(\frac{\bm{\alpha}_i\times
\bm{r}_i}{r_i^3}\right)
_z\vert\Psi_{^{2}\Sigma_{1/2}}
   \rangle, \\
\end{equation}
\begin{equation}
 \label{Aper}
A_{\perp}=\frac{\mu(\rm Ra)}{I}
   \langle
   \Psi_{^{2}\Sigma_{1/2}}\vert\sum_i\left(\frac{\bm{\alpha}_i\times
\bm{r}_i}{r_i^3}\right)
_+\vert\Psi_{^{2}\Sigma_{-1/2}}
   \rangle, \\
\end{equation}
with $\mu(\rm Ra)$ being the magnetic moment of a Ra isotope with nuclear spin quantum number $I$, and are heavily determined by the core region of the electronic wave function.

\section{Electronic structure calculations}

In this work, we are mostly concerned with quantities, which are mean values of the operators heavily concentrated in the atomic core of Ra and sensitive to variation of core-region densities of the valence electrons (the ``atom in a compound'' or AiC properties \cite{Titov:14} below). Efficient computations of AiC properties can be performed by the two-step approach \cite{Titov:06amin, Petrov:02} utilizing the generalized relativistic effective core potential (GRECP) method \cite{Titov:99, Mosyagin:10a} \footnote{It should be noted that recently there were reports on using of all-electron four-component methods for calculation of effective electric field and hyperfine structure constants, e.g.,\cite{Parpia:97,Quiney:98,Fleig:14}.}. In the first (molecular) step the GRECP is used to exclude the inner-core electrons from a correlation calculation and obtain an accurate description of the valence part of the wave function in an economical way, thus, dramatically reducing the computational cost of the relativistic molecular calculation. Second, a nonvariational restoration procedure is employed \cite{Titov:06amin} to recover the valence wave function in the inner core region of a heavy atom. The two-step approach has been used in various calculations of AiC properties \cite{Baklanov:10, Isaev:04, Skripnikov:09, Skripnikov:11a, Petrov:11, Kudashov:13, Petrov:13,Skripnikov:13c} and has proven to be a reliable source of theoretical data for experimental investigations \cite{Cossel:12}. The GRECP from Ref.~\cite{Kudashov:13} was used for Ra in this work.

\begin{table}[b]
  \caption{Equilibrium internuclear distance $R_e$ (in units of the Bohr radius   $a_0$), harmonic vibrational wavenumber $\omega_e x_e$ (in cm$^{-1}$) and vibrational anharmonicity $\omega_e x_e$ (in cm$^{-1}$) of ${}^{223}$RaF.}
  \begin{tabular}{lccc}
   \hline \hline
   Method  							& $R_e$ & $\omega_e$	& $\omega_e x_e$  \\
   \hline   
   CCSD(T) \cite{Borschevsky:13} \footnotemark[1]
   \footnotetext[1]{Two-component relativistic coupled-cluster approach with 
   single, double, and perturbative triple excitations.}
   									& 4.26  & \textemdash & \textemdash \\
   \hline
   FS-RCCSD-1 \cite{Isaev:13} \footnotemark[2]
   									& 4.24  & 428   & \textemdash \\
   FS-RCCSD-2 \cite{Isaev:13} \footnotemark[2]
   \footnotetext[2]{FS-RCCSD-1 from \cite{Isaev:13} refers to four-component Fock-space coupled cluster calculations with single and double cluster amplitudes (as implemented in DIRAC program package \cite{DIRAC12}) with Dyall's relativistic basis set and FS-RCCSD-2 with the RCC-ANO basis set by Roos et al. The active space in FS-RCCSD-1 and FS-RCCSD-2 was restricted by energy for Dyall's basis set up to 10~$E_\mathrm{h}$ (10 Hartree) and for RCC-ANO basis set up to 1000~$E_\mathrm{h}$.} 
   									& 4.29  & 431   & \textemdash \\
   \hline
   \multicolumn{4}{c}{This work} \\
   FS-RCCSD 						& 4.23  & 435	& 1.53	     \\
   \hline \hline
  \end{tabular}
  \label{spec_props}
\end{table}

Two methods were employed to treat electron correlation and relativistic effects: i) a relativistic two-component Fock-space coupled-cluster approach with single and double cluster amplitudes (FS-RCCSD) \cite{RCC:02} and ii) a spin-orbit direct configuration interaction (SODCI) approach \cite{SODCI:01, SODCI:02, SODCI:03} (modified in \cite{Titov:01} to account for spin-orbit interaction in the configuration selection procedures). 

\begin{table*}[t]
  \caption{\it Ab initio \rm calculations of AiC properties and spectroscopic constants for the $^{2}\Sigma$ ground state of RaF (the $^{223}$Ra isotope was considered in this work): P$-$odd interaction constant $\Wa$ (Hz), T,P$-$odd interaction constants $W_d$ ($\cdot 10^{25}$ $\mathrm{Hz} \cdot \mathrm{cm}^{-1} \cdot e^{-1}$), $W_\mathrm{SP}$ (kHz) and $X$ ($a_0^{-4}$); hyperfine constants $A_{||}$ (MHz) and $A_{\perp}$ (MHz), and the total angular momentum projection quantum number $J_x$.} 
  \begin{tabular}{lccccccc}
  \hline \hline
   Method  			&$\Wa$		 &$W_d$&$W_\mathrm{SP}$&X&$A_{||}$&$A_{\perp}$ &$J_x$\\ 
   \hline
   DHF \cite{Borschevsky:13} \footnotemark[1]
\footnotetext[1]{Four-component Dirac-Hartree-Fock.}
                    & 1364
                                     & \textemdash
   									   & \textemdash
   										  	 & \textemdash
   										  			  & \textemdash & \textemdash
   										  				 		   &\textemdash \\
   \hline
   GHF-ZORA \cite{Isaev:10} \footnotemark[2]
   \footnotetext[2]{Two-component generalized Hartree-Fock (GHF) in zero-order regular approximation (ZORA).}
   					& 1300		 
   					& ($-$)2.20\footnotemark[3]\footnotetext[3]{Computed via $|E_{\rm eff}/\Omega| = |W_d|$ with the value of $|E_{\rm eff}|$ estimated in Ref. \cite{Isaev:13} using the GHF-ZORA value of $W_{\rm SP}$ and the approximate ratios between $W_{\rm SP}$ and $W_d$}
 						   &$-$150\footnotemark[4]
   	 \footnotetext[4]{GHF-ZORA value from \cite{Isaev:13}.} 
   									  		& \textemdash & 1900
   									  		  & 1860
   									  		    &\textemdash \\
     GHF-ZORA scaled  \footnotemark[5]
   \footnotetext[5]{Spin-polarisation is included using scaling relations between hyperfine tensor components
   and P$-$odd properties (see \cite{Isaev:13})}
   					& 2100		 & \textemdash  & \textemdash & \textemdash & \textemdash	 & \textemdash     &\textemdash \\
   GKS-LDA \cite{Isaev:12} \footnotemark[6]
   \footnotetext[6]{Two-component density functional theory (DFT) calculations within the generalized Kohn-Sham (GKS) framework with the local-density approximation (LDA) exchange-correlation functional.}  
   					& 1470		 & \textemdash
   									  & \textemdash & \textemdash
   										  	 & \textemdash
   										  			  & \textemdash
   										  				 		   &\textemdash \\
   \hline
   \multicolumn{7}{c}{This work} \\
   SODCI            & 1540		 & $-$2.40& $-$131 & -3700 & 1790	  &1720	   & 0.491 \\
   FS-RCCS				& 1455		 & $-$2.25& $-$122 & -5620 & 1700	  &1630	   & 0.487 \\
   FS-RCCSD			& 1700		 & $-$2.65& $-$144 & -4300 & 2100	  &2020	   & 0.491\\
   CCSD				&$\mkern 8.1mu$ \textemdash \footnotemark[7] 
\footnotetext[7]{Computation of the parameter is not implemented in the current version of the code.}
     								 & $-$2.36& $-$128 & -3090 & 2090	  &$\mkern 2.1mu$ \textemdash \footnotemark[7] 
     								 							&$\mkern 2.1mu$ \textemdash \footnotemark[7] \\ 
   CCSD(T)			&\textemdash
   								 & $-$2.33& $-$127 & -3000 & 2110	  &\textemdash &\textemdash \\ 
   CCSD$_{\rm enlarged}$\footnotemark[8] \footnotetext[8]{CCSD with larger basis sets.}
   					&\textemdash
   								 & $-$2.30& $-$125 & -3140 & 2080	  &\textemdash &\textemdash \\ 
   Final			& 1700		 & $-$2.56& $-$139 & -4260 & 2110	 &2020	   & 0.491 \\ 
   \hline \hline
  \end{tabular}
  \label{core_props}
\end{table*}

The FS-RCCSD scheme begins with a one-component self-consistent-field (SCF) calculation of the reference wave function (in this case a closed-shell RaF$^+$ reference state) followed by the two-component RCCSD calculations of RaF taking account of single and double cluster amplitudes. Ten electrons of Radium 6$s^2$6$p^6$7$s^2$) and nine electrons of Fluorine (1$s^2$2$s^2$2$p^5$) were treated explicitly in the correlation calculations.

The AiC properties are calculated via the finite field method \cite{Kunik:71, Monkhorst:77}. Triple cluster amplitudes and basis set enlargement corrections for values obtained within the FS-RCCSD are computed using the scalar-relativistic {\sc CFOUR} \cite{CFOUR_full} code via interface to the nonvariational one-center restoration code developed in \cite{Skripnikov:11a}. Corrections for triple cluster amplitudes were estimated using CCSD and CCSD(T) approximations, while the basis set enlargement corrections were obtained from CCSD calculations with normal and enlarged basis sets. Final absolute values of AiC properties defined by Eqs. (\ref{matrelem}, \ref{WTP}, \ref{X}, \ref{Apar}) are obtained as 
\begin{table}[b]
  \caption{Hyperfine coupling constants $A$ (given in MHz) for $^{223}$Ra$^+$.}
  \begin{tabular}{lccc}
  \hline \hline
  \rule{0cm}{9pt}
									& $A(^{2}S_{1/2})$ & $A(^{2}P_{3/2})$ & $A(^{2}P_{1/2})$ \\ 
   \hline
   This work								& 3379			   & 64				  & 657				 \\ 
   Experiment \cite{Ra+_HFS:01, Ra+_HFS:02}	&$\mkern 17mu$ 3404(2)		   &$\mkern 17mu$ 57(8)			  &$\mkern 17mu$ 667(2)			 \\ 
   \hline \hline
  \end{tabular}
  \label{Ra+_HFS}
\end{table}
\begin{equation}
\begin{array}{l}

\rm{Y}({\rm FINAL}) = \rm{Y}({\rm FS{-}RCCSD})~ +\\[0.5em]
+ ~\left(\rm{Y}({\rm CCSD_{enlarged}}) - \rm{Y}({\rm CCSD})\right)~ +\\[0.5em]
+ ~\left(\rm{Y}({\rm CCSD(T)}) - \rm{Y}({\rm CCSD})\right).\\

\end{array}
\end{equation}

The GRECP/RCCSD method with scalar-relativistic corrections for triple cluster amplitudes and basis set enlargement was also used to calculate the ground-state potential curve of the RaF radical, which was then used to compute spectroscopic constants ($R_e$, $\omega_e$, $\omega_e x_e$) of RaF (Table \ref{spec_props}) via the Simons-Parr-Finlan potential \cite{spf:01}. HFS constants $A$ for different states of the $^{223}$Ra$^+$ ion were obtained within the FS-RCCSD method (Table \ref{Ra+_HFS}). The magnetic moment $\mu=0.271$ (in nuclear magnetons) and the nuclear spin $I=3/2$ were implied for the $^{223}$Ra nucleus.  Basis sets (20s,20p,10d,8f,5g)/[6s,8p,4d,2f,1g] \cite{Kudashov:13} and (10s,5p,2d)/[4s,3p,2d] (aug-cc-pVDZ basis set \cite{Kendall:92}) were used for Ra and F, respectively, except when computing basis set enlargement corrections, in which case basis sets (15s,15p,10d,8f,5g) and (11s,6p,3d,2f) (uncontracted aug-cc-pVTZ \cite{Kendall:92}) were used. Also in atomic calculations of $^{223}$Ra$^+$ basis set (20s,20p,10d,8f,5g)/[6s,8p,5d,5f,1g] was employed. All molecular calculations of AiC properties were carried out for the equilibrium internuclear distance, $R_e$ = 4.23~$a_0$ (2.24 \AA), the results are given in Table \ref{core_props}; the results of FS-RCCS calculations are also presented to demonstrate double cluster amplitudes' corrections.

\section{Results and discussions}

The goal of this study was to assess the possibility of using RaF in search for P$-$ and T,P$-$symmetry violation in molecules. While providing higher level of precision, our results mostly support those published recently \cite{Isaev:10, Isaev:12, Borschevsky:13, Isaev:13}, confirming RaF as a versatile multipurpose probe for fundamental symmetries violation search in low-energy sector. Triple cluster amplitudes and basis set enlargement corrections to $\Wa$ and $A_{\perp}$ cannot be obtained within the scalar-relativistic approach employed herein, because these properties require mixing of the states with different projections of the total electronic angular momentum, which is not implemented in the codes used. As for other parameters from Table \ref{core_props}, it is clear that the aforementioned corrections contribute less than 4\% to the final values. There are no obvious reasons to expect that further enlargement of the basis set and accounting for quadruples amplitudes will influence the results by more than 5\%. Taking into account our previous findings for RaO \cite{Kudashov:13}, \mbox{$X(\rm{RaF}) \approx$ $0.6 \cdot X(\rm{RaO})$}, making RaF less sensitive to T,P$-$odd effects associated with the Schiff moment than RaO. HFS constants of $^{223}$Ra$^+$ were computed to demonstrate the accuracy of our approach and, as seen from Table \ref{Ra+_HFS}, one might safely assume 10\% theoretical uncertainty of our final results (our value of $A(^{2}P_{3/2})$ is within the error margin of the experimental one).

\section*{Acknowledgement}

The authors acknowledge Saint Petersburg State University for a research grant No.~0.38.652.2013. The work of A.D.K.,\ A.N.P.,\ L.V.S.\ and A.V.T. is supported by RFBR Grant No.~13-02-01406. The work of N.S.M.\ is supported by RFBR Grant No.~13-03-01307а. L.V.S.\ is also supported by the grant of President of RF No.~5877.2014.2. The molecular calculations were partly performed at the Supercomputer ``Lomonosov''.
The work of T.A.I.\ and R.B.\ is supported by the State Initiative for the Development of Scientific and Economic Excellence (LOEWE) in the LOEWE-Focus ELCH.

\bibliographystyle{./BIBs/apsrev}

\end{document}